\documentclass[12pt]{iopart}
\usepackage{epsfig,bm}
\def\bbox#1{\hbox{\boldmath${#1}$}}

\begin{document}

\title {Does HBT Measure the Freeze-out Source Distribution?}

\author{Cheuk-Yin Wong}

\address{Physics Division, Oak Ridge National Laboratory, Oak Ridge,
TN 37831, USA}
\address{\& Department of Physics, University of Tennessee, 
Knoxville, TN 37996, USA}

\vspace*{0.2cm}
\address{E-mail: wongc@ornl.gov} 

\begin{abstract}
It is generally assumed that as a result of multiple scattering, the
source distribution measured in HBT interferometry corresponds to a
chaotic source at freeze-out.  This assumption is subject to question
as effects of multiple scattering in HBT measurements must be
investigated within a quantum-mechanical framework.  Applying the
Glauber multiple scattering theory at high energies and the optical
model at lower energies, we find that multiple scattering leads
to an effective HBT density distribution that depends on the initial
chaotic source distribution with an absorption.

\end{abstract}

\vspace*{-0.7cm}
\section{Introduction}

Recent experimental measurements of HBT correlations in relativistic
heavy-ion collisions show only relatively small changes of the
extracted longitudinal and transverse radii as a function of collision
energies, and the ratio of $R_{\rm out}/R_{\rm side} \sim 1$
\cite{Data}. The difficulties of explaining these HBT measurements 
with theoretical models, known as the ``HBT puzzles'', have been
discussed by many authors \cite{Theo,Review}. In these comparisons
with theoretical models, it is generally assumed that as a result of
multiple scattering, the source distribution measured in an HBT
measurement corresponds to a chaotic source at freeze-out, in which a
detected hadron suffers its last hadron-hadron scattering.

In a recent quantum-mechanical treatment of the multiple scattering
process using the Glauber theory at high energies and the optical
model at lower energies, it was found that the HBT interferometry does
not measure the freeze-out source distribution and the effective HBT
density distribution depends on the initial chaotic source
distribution with an absorption \cite{Won03}.  Effects of collective
flows have also been investigated \cite{Won03}.

What is the physical basis for these new insights in HBT measurements?
While the detailed arguments leading to the above results have been
presented previously in Ref.\ \cite{Won03}, it is instructive to
review here the most important features of the multiple scattering
process and the HBT interferometry that give rise to the above
unconventional viewpoints.  We shall first examine the origin of the
HBT interferometry in Section II and study next how the multiple
scattering process will affect the HBT correlation function in Section
III, using the quantum-mechanical Glauber theory of multiple
scattering \cite{Gla59}.

At the recent Quark Matter 2004 Conference, Kapusta and Li
\cite{Kap04} reported new findings which support qualitatively some of
the earlier results of Ref.\ \cite{Won03}.

\section{HBT Interferometry for Chaotic Sources}

\vspace*{-0.3cm}
We examine briefly the origin of the HBT correlation for identical
bosons (pions), as reviewed in Chapter 17 of Ref.\ \cite{Won94}.  The
HBT correlation is represented by the probability $P( {k}_1, {k}_2)$
of detecting identical bosons with 4-momenta $k_1$ and $k_2$ in
coincidence, or alternatively, by the correlation function $R( {k}_1,
{k}_2)$=$[P(k_1, k_2)/P( {k}_1)P( {k}_2)]$ -1 where $P(k)$ is the
single-particle momentum distribution.

\begin{figure}[h]
\null\vspace*{-0.4cm}
{\hspace*{3.5cm}
\includegraphics[scale=0.58]{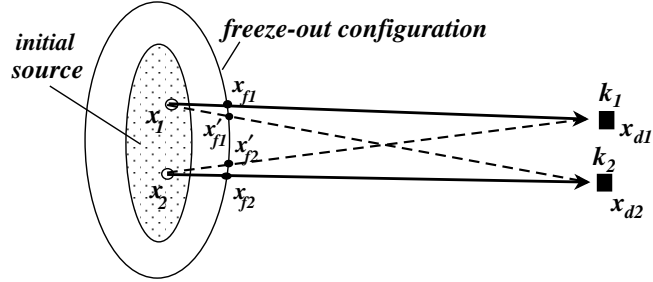}}
\vspace*{-0.4cm}
\caption{ The shaded region represents the initial source which will
\hfill \\expand to the freeze-out configuration before the bosons reach
the detectors.  }
\end{figure}
\vspace*{-0.5cm}
\noindent
$P(k_1,k_2)$ is given by squaring the total amplitude from all source
points $(x_1, x_2)$,
\begin{eqnarray}
\label{Pk1k2}
P( {k}_1, {k}_2)
={1 \over 2!}\bigl | 
\sum_{ {x}_1,  {x}_2}
A( {k}_1 {x}_1) e^{i \phi_0( {k}_1 {x}_1)}
A( {k}_2 {x}_2) e^{i \phi_0( {k}_2 {x}_2)} 
\psi_{12} ( {x}_1, {x}_2) \bigr | ^2 ,
\end{eqnarray}
where $A( {k}_i {x}_i)$ and $\phi_0( {k}_i {x}_i)$ are the amplitude
and the phase for the production of boson $k_i$ at the space-time
point $x_i$ (Fig. 1).  $\psi_{12} ( {x}_1, {x}_2)$ is the amplitude
for the propagation of a pair of bosons $(k_1,k_2)$ from the source
points $(x_1,x_2)$ to the detection points $(x_{d1}, x_{d2})$,
\begin{eqnarray}
\label{psi12}
\psi_{12}(x_1,x_2) = {1 \over \sqrt{2}}
\biggl \{ 
e^{{i  {k}_1 \cdot (  {x}_{d1}- {x}_1  )
          +i  {k}_2 \cdot (  {x}_{d2} - {x}_2  )}}
+ e^{{i  {k}_1 
        \cdot (   {x}_{d1} - {x}_2  )
          +i  {k}_2 \cdot (   {x}_{d2} - {x}_1  )}}
\biggr \},
\end{eqnarray} 
where the two terms represent amplitudes for two different histories
of traveling from the source points to the detection points (the solid
and the dashed trajectories (histories) of Fig. 1).  If the source is
coherent, there is no HBT correlation and $R( {k}_1, {k}_2)=0$.

If the source is chaotic with random and fluctuating phases $\phi_0(
{k} {x})$, the absolute square of the sum in Eq.\ (\ref{Pk1k2})
becomes the sum of absolute squares,
\begin{eqnarray}
\label{eq3}
 P( {k}_1, {k}_2)
= \sum_{ {x}_1,  {x}_2}
A^2( {k}_1 {x}_1) 
 A^2( {k}_2 {x}_2)
|\psi_{12} ( {x}_1, {x}_2) |^2.
\end{eqnarray}
The HBT correlation is then present with a non-vanishing $R( {k}_1,
{k}_2)$.  The correlation function $R(k_1,k_2)$ depends on the cross
term of $|\psi_{12}(x_1,x_2)|^2$ of Eq.\ (\ref{psi12}), which contains
the phase difference between the two histories,
\begin{eqnarray}
\label{dif}
& &
{ {k}_1 \cdot (   {x}_{d1} -  {x}_1  )
   + {k}_2 \cdot (   {x}_{d2} -  {x}_2 )}
{ -\left \{  {k}_1 \cdot (   {x}_{d1} -  {x}_2  )
   + {k}_2 \cdot (   {x}_{d2} -  {x}_1 ) \right \}}
\nonumber\\
& &=-( {k}_1- {k}_2)\cdot {x}_1 - ( {k}_2- {k}_1)\cdot {x}_2.
\end{eqnarray} 
Representing  $\sum_{x_1}$ by $\int d {x}_1\rho( {x}_1)$ with the
source spatial density $\rho(x_1)$, Eqs.\ (\ref{psi12})-(\ref{dif})
give
\begin{eqnarray}
\label{Rk1k2}
R( {k}_1,  {k}_2) 
\sim \left |~ \, \int d {x}~~e^{-i( {k}_1 -  {k}_2)\cdot  {x}} 
~~\rho_{\rm eff} (x; {k}_1, {k}_2)~ \right |^2, 
\end{eqnarray}

\vspace*{-0.5cm}
\begin{eqnarray}
\label{reff}
\rho_{\rm eff} (x; {k}_1, {k}_2)
= {\rho( {x})
A( {k}_1,  {x}) A( {k}_2,  {x}) \over \sqrt{
P( {k}_1) P( {k}_2)}}.
\end{eqnarray}
The measurement of $R(k_1, k_2)$ therefore provides information on the
Fourier transform of the effective source distribution $\rho_{\rm eff}
(x; {k}_1, {k}_2)$ of a chaotic source.

\section{Multiple Scattering and HBT}

\vspace*{-0.3cm} In high-energy heavy-ion collisions, particles such
as pions are produced and they will undergo multiple scattering until
the system reaches the freeze-out configuration (Fig. 1).  According
to the Glauber theory of multiple scattering \cite{Gla59}, the
probability amplitude for a pion with momentum $ {k}_1$ to go from the
source point $ {x}_1$ to the freeze-out point $x_{f1}$ and then to the
detection point ${x}_{d1}$ is
\begin{eqnarray}
\label{amp}
\psi( {x}_1\to {x}_{d1})= \exp \{   
    i  {k}_1 \cdot 
 ( {x}_{d1} -  {x}_1 ) +i\phi_s( {x}_{f1}- {x}_1) \} .
\end{eqnarray}
At high particle energies, the phase $\phi_s( {x}_{f1}- {x}_1)$ is the
sum of two-body scattering phase shifts $\chi_j$ of the $N$ particles
with which the particle scatters along its trajectory,
\vspace*{-0.1cm}
\begin{eqnarray}
\phi_s( {x_{f1}}- {x}_1) 
= \sum_{j}^{N(x_{f1||}-x_{1||})}
\chi_j( \bbox{x}_{1\perp}-\bbox {x}_{j\perp}) ,
\end{eqnarray}
where $x_{1||}$ and $\bbox{x}_{1\perp}$ are coordinates longitudinal
and transverse to $\bbox{k}_1$. At lower energies, $\phi_s( {x}_{f1}-
{x}_1)$ is given in terms of the two-body optical potential
$V_j(\bbox{x}_1-\bbox{x}_j)$ and the particle velocity $v$,

\vspace*{-1.0cm}
\begin{eqnarray}
\phi_s( {x}_{f1}- {x}_1) = -\int_{x_{1\|}}^{{x_{f1\|}}}  
{1 \over v}\sum_{j}^{N(x_{f1||}-x_{1||})}
V_j ({\tilde {\bbox{x}}}_1 - {\bbox{x}}_j    ) d{\tilde x}_{1\|}.
\end{eqnarray}
The probability amplitude (\ref{amp}) leads to the proper classical
transport description of the particle in a medium \cite{Won04}.  It
modifies the wave function $\psi_{12}(x_1,x_2)$ of Eq.\ (\ref{psi12})
to
\begin{eqnarray}
\psi_{12} (x_1,x_2) = {1 \over \sqrt{2}}
\bigl \{ & &
e^{{ i  {k}_1 \cdot (  {x}_{d1}- {x}_1  )+i\phi_s( {x}_{f1}- {x}_1)
          +i  {k}_2 \cdot (  {x}_{d2} - {x}_2  )
+ i \phi_s( {x}_{f2}- {x}_2)  }}
~~~~~~~~~~~~~\nonumber
\\
&+& e^{{ i  {k}_1 
        \cdot (   {x}_{d1} - {x}_2  )+i\phi_s( {x}_{f2}'- {x}_2)
          +i  {k}_2 \cdot (   {x}_{d2} - {x}_1  ) 
+i\phi_s( {x}_{f1}'- {x}_1) }}
~~~\bigr \} ,
\end{eqnarray}
where $({x}_{f1} {x}_{f2})$ and $({x}_{f1}' {x}_{f2}')$ are the two
sets of freeze-out coordinates for the two histories (Fig. 1).  In an
HBT measurement of a chaotic source, the difference of the phases in
the cross term of $|\psi_{12}(x_1,x_2)|^2$ in Eq.\ (\ref{dif}) is now
modified to
\begin{eqnarray}
\label{dif2}
& &{~~~~   {k}_1 \cdot 
(   {x}_{d1} -  {x}_1 ) +\phi_s( {x}_{f1}- {x}_1)
 \,  + {k}_2 \cdot (   {x}_{d2}-  {x}_2  )
+\phi_s( {x}_{f2}- {x}_2)} 
\nonumber \\
& &{ - \{  {k}_1 
  \cdot (   {x}_{d1} -  {x}_2 )+\phi_s^*( {x}_{f2}'- {x}_2)
   + {k}_2 \cdot (   {x}_{d2} -  {x}_1  )
   +\phi_s^*( {x}_{f1}'- {x}_1)
\}}  \nonumber\\
& &~~~~~ \equiv -( {k}_1- {k}_2)\cdot {x}_1 -
( {k}_2- {k}_1)\cdot {x}_2 +  \Delta,
\end{eqnarray} 
\noindent
where we introduce $\Delta$ to represent the effects of multiple
scattering. For the measurement of $R_{\rm out}$, $ \bbox{k}_1-
\bbox{k}_2$ is along $
\bbox {k}_1$, we have $ \bbox{x}_{f1}=\bbox{x}_{f1}' {\rm
~and~} \bbox{x}_{f2}=\bbox{x}_{f2}'$, 
\begin{eqnarray}
 \phi_s( {x}_{f1}- {x}_1) = \phi_s( {x}_{f1}'-
{x}_1),{~~\rm and~~} 
\phi_s( {x}_{f2}- {x}_2) = 
\phi_s({x}_{f2}'- {x}_2).
\end{eqnarray}
For the measurement of $R_{\rm side}$ and $R_{\rm long}$,
$\bbox{k}_1-\bbox{k}_2=\bbox{q}$ is perpendicular to $\bbox{k}_1$, and
\begin{eqnarray}
\fl
~~~~~~~~~~~
{\cal R}e~\phi_s( {x}_{f1}- {x}_1)  
  -{\cal R}e ~\phi_s( {x}_{f1}'- {x}_1)
&=& 
( {x}_{f1}- {x}_{f1}') \cdot
\nabla {\cal R}e ~\phi_s( {x}_{f1} - {x}_1)
\nonumber\\
&\propto&~~~ \bbox{q} \cdot  \nabla_{\perp} 
{\cal R}e ~\phi_s( {x}_{f1}- {x}_1)
~~\sim~ 0, 
\end{eqnarray}
where the last approximate equality arises as the vector sum in
$\nabla_{\perp} {\cal R}e ~\phi_s$ with random transverse vector
contributions from many independent scatterings is approximately zero.
In both cases, the real parts of the phase differences cancel
approximately, and only the imaginargy parts remain,
\begin{eqnarray}
\label{del}
\Delta \sim  2i~{\cal I}m ~\phi_s( {x}_{f1}- {x}_1) 
            +2i~{\cal I}m ~\phi_s( {x}_{f2}- {x}_2).
\end{eqnarray}

\vspace*{-0.1cm}
\noindent
From Eqs.\ (\ref{dif2}) and (\ref{del}), $R(k_1,k_2)$ is again given
by Eq.\ (\ref{Rk1k2}) but the HBT effective density distribution
$\rho_{\rm eff}(x;k_1,k_2)$ of Eq.\ (\ref{reff}) is modified, in the
case of multiple scattering, to \cite{Won03}
\begin{eqnarray}
\label{abs}
\rho_{\rm eff} (x; {k}_1, {k}_2)= { e^{-2 ~{\cal I} {m}~ \phi_s( {x}_f-x)} 
\rho( {x})
A( {k}_1,  {x}) A( {k}_2,  {x}) \over \sqrt{
P( {k}_1) P( {k}_2) } } \,.  
\end{eqnarray}
The effective distribution depends on the initial source distribution
with an absorption.

\vspace*{-0.3cm}
\section{Conclusions and Discussions}

\vspace*{-0.3cm}
Comparing the results of the last two sections, we see that the new
insights concerning the multiple scattering process and HBT
measurements arise from the following considerations: (1) In a quantum
mechanical description, the multiple scattering process gives rise to
the accumulation of phases along the trajectories of the detected
particles. (2) For a chaotic source, the HBT correlation arises from
the difference in the phases accumulated in two different sets of
particle trajectories.  (3) For these two sets of trajectories, the
real parts of the accumulated phases due to the multiple scattering
process approximately cancel each other and only the imaginary
absorptive parts remain.

Based on these considerations, we find that the multiple scattering
process leads to an effective density distribution that depends on the
initial chaotic source distribution with an absorption, Eq.\ (\ref{abs}).
The effects of the longitudinal momentum loss can be further included in
the future by using the extension of the Glauber theory formulated by
Blankenbecler and Drell \cite{Bla96}.  While the absorption and the
longitudinal momentum loss will modify the transmission of the initial
chaotic source distribution, the effective source is closer to the
initial chaotic source configuration and the nuclear geometrical
overlap than the freeze-out configuration.  As a consequence, the
present new insights may pave the way for a better understanding of
the HBT puzzles.

\vspace*{+0.3cm}
\noindent
{\bf Acknowledgment}

\vspace*{-0.0cm}
\noindent 
The author wishes to thank Profs.\ R.\ Blankenbecler, R.\ Glauber, and
Weining Zhang for valuable discussions. This research was supported by
the Division of Nuclear Physics, U.S. D.O.E., under Contract
DE-AC05-00OR22725 managed by UT-Battelle, LLC.

\end{document}